\documentclass[aps,prd]{revtex4}

\usepackage{psfig}
\newcommand{\beq}{\begin{equation}}
\newcommand{\eeq}{\end{equation}}
\newcommand{\beqa}{\begin{eqnarray}}
\newcommand{\eeqa}{\end{eqnarray}}
\newcommand{\beqar}{\begin{eqnarray*}}
\newcommand{\eeqar}{\end{eqnarray*}}

\def \s {\,\,\,\,}

\def \sk {\nonumber\\}
\def \d {\delta}

\def \bh {black hole}
\def \Sch {Schwarzschild}

\def \a {a_o}
\def \wo {\omega_o}
\def \wqnm {\omega_{QN}}
\def \qnm {QNM}
\def \qnms {QNM's}
\def \lett {letter}
\def \gn {\gamma_{n\d}}

\newcommand{\G[1]} {\Gamma_{#1\d}}
\def \p {p_{n, \d'}}
\def \pd {p_{n,\d}}
\def \po {p_{n,1}}
\def \k {k}
\def \bi {\gamma}
\def \lp {l^2_P}
\def \jm {j_{min}}
\def \lqg {LQG}
\begin{document}


\title{The spectrum of quantum black holes and quasinormal modes}


\author{Jonathan Oppenheim
  }

\affiliation{
Racah Institute of Theoretical Physics, Hebrew University of Jerusalem, 
Givat Ram, Jerusalem 91904, Israel}


\date{July 11th, 2003}
\begin{abstract}
The spectrum of multiple level transitions of the quantum \bh\
is considered,  and the line widths calculated.  Initial evidence
is found for these higher order 
transitions in the spectrum of quasinormal modes
for \Sch\ and Kerr \bh s, further bolstering the idea that there exists
a correspondence principle between quantum transitions 
and classical ``ringing modes''.  Several puzzles are noted, including
a fine-tuning problem between the line width and the level degeneracy.
A more general explanation is provided for why setting the 
Immirzi parameter of loop
quantum gravity from the \bh\ spectrum
necessarily gives the correct value for the \bh\ entropy.
\end{abstract}



\maketitle

Although there is a lack of experimental data on which to base attempts
to construct a quantum theory of gravity, it is commonly regarded
that the theory must give a correct accounting for the \bh\ entropy.
The fact that the \bh\ entropy is a quarter of the \bh\ area then
plays the role of an experimental data point on which to test any theory.  
String theory gives the correct prediction for extremal black holes
\cite{strominger-bhe},
and loop quantum gravity gives the entropy up to a proportionality
constant \cite{ashtekar-bhe,rovelli-bhe} 
known as the Immirzi parameter.  In general, the 
area-scaling of entropy is rather generic for gravitating 
systems \cite{area}.

A less ambitious program involves 
attempting to quantize the \bh.
As early as 1974, Bekenstein \cite{bek-qbh,bek-qbh-review}, 
made the case that the area $A$ of the quantum \bh\ is quantized 
with equal spacing between levels 
\beq
A=\a n \s\s\s n=1,2,3...
\label{eq:equala}
\eeq 
in units where $G=\hbar=k_B=1$ and $a_o$ a constant \cite{nnotqnm}.
This has been the standard starting point for the quantum black
hole 
\cite{muk-qbh,qbh-others}, 
as it is based on general arguments
rather than on a particular model.  
Bekenstein
argued that since classically the 
black hole's area is an adiabatic invariant, it should be quantized
(following an insight of Ehrenfest \cite{ehrenfest-ad}).
Furthermore, for non-extremal black holes, he argued that the minimum 
change in area is independent of the black hole mass, charge and
angular momentum, which naturally leads to Eq. (\ref{eq:equala}).

Even in his original paper, 
(also \cite{bek-qbh-review}) Bekenstein
noted that Bohr's correspondence principle
implies that transitions in energy level of the quantum black
hole correspond to the black hole's quasinormal 
``ringing modes'' (\qnms)
\cite{qnm-review}.  For large $n$, one expects the quantum \bh\ to
correspond to the classical \bh\ just as a quantized oscillator
in the large mass limit
should give the correct normal modes of a classical oscillator.
Since the mass $M$ of a \bh\ 
is given by 
$\sqrt{A/16\pi}$, the energy $\wo=\Delta M$
emitted when the \bh\ looses one
area quantum is given by
\beq
\wo=\frac{\a}{32\pi M} \s .
\label{eq:wogen}
\eeq 

It was noted by Bekenstein and Mukhanov 
\cite{muk-qbh,bek-muk-95}, that the
constant $\a$ should be $4$ times the logarithm of a natural number if one
is to interpret the quantum levels of the \bh\ as giving rise to
the Bekenstein-Hawking entropy $S$.  
\beq
S=\frac{1}{4} A
\label{eq:bhentropy}
\eeq
We can use the fact that that the entropy of a \bh\ is $\ln g_n$
where $g_n$ is the number of states at level $n$.  Defining the
ground state degeneracy as $\k=g_1$, we can use Eqs.
(\ref{eq:equala}) and (\ref{eq:bhentropy}) to fix 
$\a=4 \ln{\k}$ with $\k$ a natural number.  
\beq
\wo=\frac{\ln \k}{8\pi M} \s .
\label{eq:wo}
\eeq

Hod \cite{hod1}, then noticed that the \qnm\ spectrum for
the \Sch\ \bh\ had a frequency whose real part numerically approached 
Eq. (\ref{eq:wo}) with $k=3$ in the limit of infinite imaginary frequency).
Motl \cite{motl1} later confirmed this analytically. In light of this,
Dreyer \cite{dreyer1} proposed changing the gauge group of loop quantum
gravity from SU(2) to SO(3).  He then
advocated 
using the spacing of quasinormal modes to fix the undetermined
Immirzi parameter.  He argued that the value that fixes
the energy spacing also yields the correct value for the \bh\ entropy,
thus claiming the \bh\ entropy a prediction of the theory. 
This has generated a great level of excitement in the field, and 
since then, a large number of studies have been conducted both to 
extend our understanding of quasinormal modes 
\cite{qn-experimental}
and to further understand the quantum \bh\ in this
context \cite{qn-theory}.

To learn about quantum \bh s by studying the \qnm\
structure of classical \bh s is certainly a speculative
undertaking.  Nonetheless, given the highly intriguing numerical 
coincidences
which are emerging, 
and the lack of real experimental data on which
to base a quantum theory of gravity, 
there is merit in taking 
the preceding arguments seriously, and seeing how far they can be 
pushed.  
Certainly a study of the phonon modes of a solid would
give one insight into their quantization.  Whether the \qnm\ spectrum
of the \bh\ can be treated in the same way as Bohr treated
experimental data from
the hydrogen atom 
remains to be seen.  In the remainder of the paper, I will
essentially assume that such a correspondence holds.
 
Thus far, researchers have only looked for transitions
involving one area quantum and focused on the \qnm\ spectrum
in the limit of large damping.  No convincing explanation exists as
to why this should be the significant regime, although some
suggestions have been made
(e.g. \cite{hod-kerr}).  If one believes that the quasinormal modes
arise from the underlying quantum structure, then one expects that
this correspondence principle 
should apply to all modes.
There does not appear to be any reason
to single out the highly damped modes as arising from the quantum
structure, and the rest of the modes as arising from some other
structure.  One is therefore forced to look for an explaination
for the less damped mode as well.

Furthermore, there has been no attempt to link the
imaginary part of the spectrum with the quantum \bh.
Motl \cite{motl1} has noted that the spacing of the imaginary 
part corresponds to the expected poles in the thermal Green's function.
It is unclear why this correspondence also only appears at large
damping nor how it arises from the quantum structure of the \bh.

If one takes Bohr's correspondence principle seriously, it is natural that
the line width
of the quantum \bh\ be associated with the imaginary part of the \qnms\
(since one expects classical damping or dispersion to correspond to 
the line broadening of the quantum transition
\cite{decay-review}).
I will thus first reexamine the 
quantum black hole and calculate the
line-broadening for multiple level transitions. 
One other observation is that one expects not
only transitions in which the \bh\ jumps one level,
but also higher level transitions in which the \bh\ jumps $\d$ levels.  
Then, taking the correspondence
between \qnms\ and quantum \bh s seriously
one expects to see \qnms\ with a real frequency of $\d\wo$.
Indeed, I then find that the \qnm\ spectrum
contains some evidence for multiple level transitions in addition to
the single level transition so far observed. I will present data
from both \Sch\ and Kerr, which, although not as clean as the
data in the asymptotic regime, shows initial evidence for these multiple level
transitions.  With regard to the imaginary part of the \qnm\ spectrum,
the expected scaling is observed for the multiple level
transitions, but several puzzles remain.

After presenting the data, I return to some theoretical aspects of
the quantum black hole, and note a fine-tuning problem which
exists in the physics governing the line-broadening of the spectrum.
I then discuss a puzzle, particularly if $k=3$, concerning
suppression of Hawking radiation.  Finally, I note that there is
a general
explanation (in terms of the Bekenstein model) for why 
fixing the Immirzi parameter from the
quantum \bh\ spectrum necessarily gives the correct result for
the \bh\ entropy.

Let us consider a spontaneous emission process in which a \bh\ with
$n$ area-quanta decays $\d$ levels.  If the probability per unit time of a 
spontaneous decay between two levels $n$ and $n-\d'$ is $\p$,
then under the assumption that these transitions give rise
to the thermal character of \bh\ radiation, one expects
\beq
\frac{\p}{\pd} =
e^{\beta (\d-\d')\wo} \left(\frac{\d'}{\d}\right)^2
\eeq
where $\beta$ is the inverse Hawking temperature
\beq
\beta = 8\pi M  
\eeq
and the factor $ ({\d'}/\d)^2$ comes from the phase space 
(e.g. $(\d\wo)^2$) of the emitted radiation.  One also gets
such a probability distribution if one assumes that the
degeneracy $g_n$ is what dominates the transition.
We can then use Eq. (\ref{eq:wo}) to write
\beq
\p=k^{(1-\d')}\d'^2 \po\s .
\label{eq:p1}
\eeq
Then the total probability $\Gamma_n$ per unit time
for the decay of the n'th level is
\beqa
\G[n] &=& \sum_{\d'=1}^n \p \sk
&=& \frac{\po k^2(1+k)}{(k-1)^3} + O(n^2 k^{-n}) 
\eeqa  
where we henceforth drop terms which
are exponentially suppressed for large $n$.  
Using the methods of Weisskopf and
Wigner\cite{ww-linewidth1,ww-linewidth1,decay-review},
the line-width $\gn$  
of the transition from $n$
to $n-\d$ 
is given by 
\beqa
\gn&=&\Gamma_n + \Gamma_{n-\d} \sk
&=&  \frac{k^2(1+k)}{(k-1)^3} (\po +p_{n-\d,1}) \s .
\label{eq:impart}
\eeqa
while the difference between two line-widths of a \bh\ with
fixed $n$ is
\beq
\gn-\gamma_{n,\delta'}
=
\frac{k^2(1+k)}{(k-1)^3} (p_{n-\d,1}-p_{n-\d',1)})
\label{eq:dif}
\eeq

From the Stefan-Boltzmann law and the Hawking temperature,
one can see that the classical \bh\ evaporates its mass at
a rate proportional to $1/M^2$ (i.e. $1/n$) \cite{page-rates}.
Bekenstein and Mukhanov \cite{bek-muk-95} have calculated
the probability distribution of the quantum \bh\ to make
various transitions.  They use the classical result 
to fix the decay rate of single level transitions (which dominate).  
Were we to follow this reasoning,
we could set the total luminosity to the classical result
\beq
\wo \sum_\d p_{n\d} \d \propto 1/n
\eeq 
where a constant of proportionality accounts for the non-continuous 
nature of the spectrum and  would depend on the particular
particle being emitted (with different values of the proportionally
constant being advocated by different authors\cite{makela-96,bek-muk-95,hod-broad}) 
%

Using Eq. (\ref{eq:p1}), this gives
\beq
\po= \frac{(k-1)^4}{k^2(1+4k+k^2)}\frac{b}{\sqrt{n\ln{k}}}
\label{eq:semiclass}
\eeq
with $b$ being the constant of proportionality. This can now be 
substituted into Eq. (\ref{eq:impart}) to give the line-breadth
of \bh\ transitions.

Now that we have the line-breadth and energy spacing for multiple
level transitions, let us turn to whether they 
are reflected in the \qnm\ spectrum.
%
%
\begin{figure}
\psfig{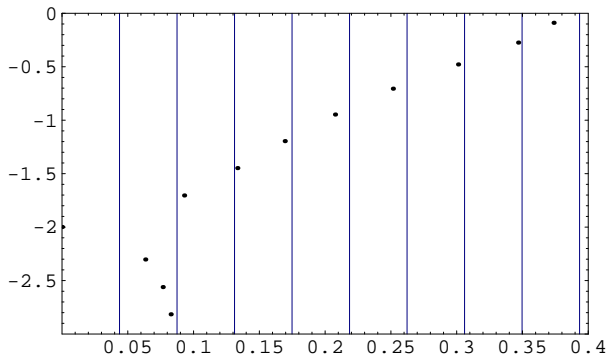}
\caption[l]{$Im(M\wqnm)$ vs. $Re(M\wqnm)$ (abscissa) for the \Sch, $l=2$  
  quasinormal modes \cite{leaver}.  The verticle
  lines are the theoretically predicted values  $M\d\wo$.}
\label{fig:l2}
\end{figure}
%
%
%
%
%
%
Fig. \ref{fig:l2} shows the gravitational perturbation with
lowest angular momentum $l=2$.  In addition to the level at
infinite damping ($\d=1$), one sees evidence for multiple transitions
which occur close to the predicted values $\d\wo$ (for $\d=2...9$).
The \qnms\ which lie close to these theoretical predictions are
the $n=1..8$ \qnm which have increasingly larger imaginary part.
The agreement with the theoretically predicted 
result is within 5\%. Strongest disagreement occurs at the highest
energy transition.  The reason that the spectrum ends here
is explained by the fact that one does not have
$\d\wo > 1/3\sqrt{3}M$ (the peak of the \bh\ potential)
since this is when the energy of the mode is larger than the peak
of the \bh\ potential.  At this energy, the mode must
become either purely outgoing or purely ingoing (while \qnms\ are defined
to be outgoing at infinity, and ``ingoing'' at the horizon
i.e. falling into the black hole at the horizon). 

The $n=9$ \qnm\ 
is the  ``algebraically special''
mode at $Re(\omega)=0$).  
Whether the latter mode is in fact a \qnm\ is
a matter of some debate \cite{qnm-review}.
Following this mode, the spectrum $n=10,11,12...$
gradually asymptots to the $\d=1$ line.   

Since the imaginary part of the \qnm\ should correspond to
the line width of the quantum \bh, one expects the imaginary
part of the \qnm\ spectrum to be given by Eq. (\ref{eq:impart}).
It is perhaps encouraging
that the higher order transitions do have an imaginary
part which is proportional to $1/\sqrt{n}$ as
Eq. (\ref{eq:semiclass}) predicts.
However, the spacing of $Im(\wqnm)$ between different modes
does not 
correspond to Eq.  (\ref{eq:dif}).  
The entire \qnm\ spectrum scales like $1/M$ (i.e. $1/\sqrt{n}$).  
From Eqs. (\ref{eq:impart}) and (\ref{eq:dif}) we see that
if $\gn$ scales like $1/M$, then the difference in line-width
between two successive transitions of different $\d$ should
essentially be the derivative of $\gn$, and one would expect
$Im(\wqnm)$ to have a term which scales like $-\d /M^2$.
This is not observed, although the slope
of $Im(\wqnm)$ does go in the expected direction in that 
higher order transitions are sharper.  
There are sets of modes which occur at roughly equal
$Im(\wqnm)$ as in (\ref{eq:impart}), but these are at higher $l$.  

The behavior of the \qnm's which would correspond to multiple level 
transitions are in stark contrast
to the \qnm\ meant to correspond to the transition at $\wo$.
The latter is 
infinitely broad (occurring at
infinite imaginary frequency), and  surrounded by a 
huge degeneracy of other modes.
While one can find many possible explanations for the
splitting of the energy levels, or to explain why particular transitions
should be broad or narrow,  I know of no general arguments which could 
consistently and convincingly explain the different behavior (witnessed
in the \qnm\ spectrum) between
the first level transition and the multiple level ones.

While the $\d=1$ \qnm\ corresponds exactly to $\wo$, the data
for $\d>1$ is not exact.  This could be for a number of reasons.
One expects the energy levels of the \bh\ to be shifted
because of their coupling to fields.  Additionally, it is not at all clear
the extent to which \qnms are probing the structure of the 
\bh\ horizon.  The considerations here are at best an approximation
to the actual quantum structure of the \bh.

Preliminary analysis of perturbations of higher $l$ show 
mixed results.  The \qnm\ data for Kerr, initially 
calculated by Leaver \cite{leaver}, also is less clear, having
a very rich structure.  Data
from Ref. \cite{bcko-kerr}, is plotting in Fig. \ref{fig:kerr15} and
Fig. \ref{fig:kerr2}.  The theoretical prediction
\cite{bek-qbh,bek-qbh-review}
\beq
\d\wo = T_H \ln{k}\, \d + 4\pi J/(MA) m
\eeq
(with $T_H$ the Hawking temperature, and J the \bh's angular momentum)
is also shown. 
Thus far, researchers
have focused their attention on modes in the asymptotic regime, 
thus concluding that the Kerr spectrum does not show
evidence for  \bh\ quanta (since only the 
$4\pi J/(MA) m$ 
term  is found in this limit) \cite{bcko-kerr,hod-kerr}. 
However, by taking into consideration the non-asymptotic part of the
\qnm\ spectrum, I would argue that the
behavior of the quasinormal modes of Kerr 
suggests the existence of transitions where the \bh\ area changes
by some number of quanta.  
%

%

One does observe fairly equal spacing as theoretically 
predicted, although (for example) the levels of $J/M=.2$ occur at 
the half-tones. 
The behavior of these plots is fairly typical, with a set of
modes at gradually sloping line-broadening (imaginary part), followed
by a sudden (and remarkable) change at the $\d=1$ level.  
The spectrum then gradually
asymptotes from the $\d=1$ level to the $\d=0$ level.

Let us now turn to two theoretical aspects of the quantum
\bh\ which do not depend on the QNM spectrum but on which the 
QNM spectrum might shed light.  First,
there is an interesting fine-tuning problem with expression
(\ref{eq:semiclass}) which is worth noting.  As explained in the
discussion preceding this equation, if one wants $\po$
to agree with the classical result then we require that it scale
like $1/\sqrt{n}$.  On the other hand, if we assume that we can
apply Fermi's Golden rule to the \bh\ (this only requires
that the decay is governed by some transition Hamiltonian), 
then we seem to get
different behavior.  Namely, 
\beqa
\po & = & (2\pi)^2 \wo^2 g_n T_{n,n-1}^2  \sk
&=& \frac{4\pi k^n \ln{k}}{n } T_{n,n-1}^2
\label{eq:goldenrule}
\eeqa
Where $T_{n,n-1}$ is the transition matrix from the $n$'th \bh\ state
to the $n-1$ state. 
Since presumably, the strength $T_{n,n-1}$ of the transition matrix 
and the degeneracy $g_n$ of the levels are independent of each other,
it is rather surprising that they should conspire in precisely the
exact way to give Eq. (\ref{eq:semiclass}).  One possibility is
that something like Eq. (\ref{eq:goldenrule}) is correct, and
that some other processes are involved in determining the
classical emission rate.  This would then explain the fact that
the real part of the \qnm\ spectrum approaches $\wo$ only in
the limit of infinite damping, since here, the absorption
time is exponentially fast in $n$.  Such an explanation however,
would not explain why the higher level transitions do have
a lifetime of $1/\sqrt{n}$ according to the \qnm\ spectrum.
There is a natural model to circumvent this problem: 
assume 
%
%
%
%
%
\begin{figure}
\psfig{file=kerr075.epsi,width=8cm}
\caption[l]{$Im(M\wqnm)$ vs. $Re(M\wqnm)$ (abscissa) for Kerr $J/M=.15$, $l=m=2$  
  \cite{bcko-kerr}.}
\label{fig:kerr15}
\end{figure}
\begin{figure}
\psfig{file=kerr1.epsi,width=8cm}
\caption[l]{$Im(M\wqnm)$ vs. $Re(M\wqnm)$ (abscissa) for Kerr $J/M=.2$, $l=m=2$  
  \cite{bcko-kerr}}
\label{fig:kerr2}
\end{figure}
\begin{figure}
\psfig{file=kerr25.epsi,width=8cm}
\caption[l]{$Im(M\wqnm)$ vs. $Re(M\wqnm)$ for Kerr $J/M=.5$, $l=m=2$  
  \cite{bcko-kerr}}
\label{fig:kerr5}
\end{figure}
%
%
%
%
the decay 
is dominated
by transitions with little change in the degrees of freedom associated
with each quanta.  If we label the $k$ degrees of freedom of each
area quanta by $s_i$, then it is rather natural to regard a transition as
the disappearance of a single quanta where none of the other quanta
change $s_i$. I.e. the other area quanta remain as passive observers
of the transition.  This leads to an effective degeneracy of the transition
of $n$ rather than $k^n$, since this process can occur by any of the
quanta being annihilated.  This would not effect the entropy, since
there are still $k^n$ possible states.  It is arguably also more simple
than the Bekenstein-Mukhanov transition, since only one quanta is
involved in each decay, rather than a large collective process which
involves the entire $n$ quanta.   $T_{n(n-1)}^2$ would then just have 
to behave like $1/M$, which rather naturally occurs in simple 
harmonic-oscillator
type transitions.  However, this model has the disadvantage that it
is harder to explain the thermal character of the emitted radiation.

Another interesting puzzle worth pointing out, 
puts into question the thermal character of the radiation of the 
quantum \bh.  If $k=3$ (as is popularly supposed), then
the thermal emission of the classical \bh\ will be substantially 
suppressed.  This is because the smallest possible emission 
(corresponding to $\wo$) occurs at an energy almost identical
to $1/\beta$.  The Hawking radiation of this quanta is therefore
suppressed by an amount $1/e$.  Higher level transitions such as
those we have discussed, will be exponentially suppressed.  
Most of the Hawking emission will therefore occur at a single
frequency.  This also occurs for $k=2$ although to a lesser
extent.  The fact that these higher level transitions are so
weak, lying outside the peak of the thermal spectrum
might play a role in explaining the difference between these levels,
which are sharp,
and the huge degeneracy of broad levels which occur at large 
imaginary part.  
This is in addition to the well known issue that
the Hawking spectrum is continuous while the Bekenstein model
gives a descrete spectrum.

Finally, we address Dreyer's proposal to change the gauge group of 
loop quantum gravity (\lqg) from SU(2) to SO(3) in light of the 
\qnm\ spectrum.  
The proposal is to fix the Immirzi
parameter $\bi$ using $\wo$.  This is viewed as giving an
independent way to fix $\bi$ (instead of using the \bh\ entropy), and 
therefore the fact that it also gives
the correct value for the \bh\ entropy is viewed as a prediction
of the theory.  We now give an explanation for this
which relies on rather general arguments.  I suggest that
the ambiguity of $\bi$ still remains, but the theory does
become more testable.

First, we note that Dreyer's arguments do not depend on
the details of \lqg, except so far as \lqg\
is believed to be consistent with the Bekenstein model.
Namely, i) \lqg\ gives  Eq. (\ref{eq:equala}) with 
$a_o=8\pi\lp \bi\sqrt{\jm(\jm+1)}$
and $\jm$ the
minimum allowed spin of the spin-network edges which
puncture the surface of the horizon (although initial conclusions were
that the area spectrum was not evenly spaced \cite{lqg-uneven}).  ii) In \lqg\ each area 
quanta contributes a set amount of entropy $\ln{k}$ (with
$k=(2\jm+1)$).  iii) Dreyer assumes that
\bh\ emmission is given by the disappearance of one of these
punctures (i.e., a decrease in $n$).  These are precisely the
same conditions that gave rise to Eq. (\ref{eq:wogen}).  

The Bekenstein model has two undetermined parameters, $a_o$, and
$k$ which one can fix by setting $a_o$ to match the \bh\ entropy
i.e. $a_o=4\log{k}$, and then perhaps fixing $k$ from the \qnm\ spectrum.
Likewise, in \lqg, one can first set $\bi$ to give  $a_o=4\log{k}$ (as
was previously done, although for a fixed $k$),
and then set $k=3$ to match the \qnm\ data.  Here, one sees that
\lqg\ has two undetermined parameters which must be set to the data.
This way of setting the parameters is physically equivalent to
Dreyers' method, just the order is reversed.
 
However, what makes Dreyer's result very interesting is that
it does provide a potential test for \lqg\ -- namely the extent to which
$k$ can arise naturally.  It might
have been that one could not have $3$ degrees of freedom per puncture,
thus one hurdle has already been cleared.
Although a number of hurdles remain, strong arguments in favour of 
$k=3$ would provide
a boost to \lqg.  Presumably, $\jm$ is more tightly constrained
then $\bi$, making the prospect for constraining the theory in this
regard  brighter.

Furthermore, the fact that the \qnm\ spectrum seems to fit
with Bekenstein's prediction, supports the equal area spacing
model, and indeed, any quantum theory of gravity which gives rise to the same spectrum
(such as \lqg\ with assumption iii).
The fact that one finds  some evidence for multiple 
level transitions  further bolsters this contention, although
the evidence is not unambiguous.
Certainly one should retain a degree of healthy skepticism
about the project of making predictions using the \qnm\ spectrum. 
A number of puzzles still remain, and
regardless of the \qnm\ spectrum, we have seen that there are many
open questions concerning the
quantization of the \bh\ which can perhaps serve as a guide in
constructing a quantum theory of gravity.

\vskip .2 cm
\begin{acknowledgments}
I thank Jacob Bekenstein for many 
helpful and pleasurable discussions on the issues contained in
this \lett, as well as for comments on the draft version. Comments
by Olaf Dreyer were also greatly appreciated, as were
interesting discussions
with Ryszard and Michal Horodecki and
John Preskill. 
I am also grateful to the authors of Ref. \cite{bcko-kerr}  
for generously providing the data from their work. 
I acknowledge the support of the Lady Davis Trust, and
ISF grant 129/00-1.

\end{acknowledgments}


\begin{thebibliography}{26}
\bibitem[{nno()}]{nnotqnm}
\bibinfo{note}{$n$ should not be confused with the order of the quasinormal
  modes}.
\end{thebibliography}

\end{document}